\documentclass[aps,prb,twocolumn,showpacs]{revtex4}
\usepackage{graphicx}

\begin{document}

\title{Magnetic and Transport Properties of Fe-Ag granular multilayers}

\author{M.~Csontos$^{1}$, J.~Balogh$^{2}$, D.~Kapt\'as$^{2}$, L. F.~Kiss$^{2}$ and G.~Mih\'aly$^{1}$}
\affiliation{$^{1}$Department of Physics, Budapest University of
Technology and Economics and \\
"Electron Transport in Solids" Research Group of the Hungarian
Academy of Sciences, 1111 Budapest, Hungary\\
$^{2}$Research Institute for Solid State Physics and Optics, 1525
Budapest, Hungary}

\date{\today}

\begin{abstract}
Results of magnetization, magnetotransport and M\"ossbauer
spectroscopy measurements of sequentially evaporated Fe-Ag
granular composites are presented. The strong magnetic scattering
of the conduction electrons is reflected in the sublinear
temperature dependence of the resistance and in the large negative
magnetoresistance. The simultaneous analysis of the magnetic
properties and the transport behavior suggests a bimodal grain
size distribution. A detailed quantitative description of the
unusual features observed in the transport properties is given.
\end{abstract}

\pacs{75.47.De; 75.70.Cn; 75.20.En; 73.43.Qt}

\maketitle

\section{INTRODUCTION}
As promising candidates for magnetic recording and sensor
applications heterogeneous magnetic materials, i.~e. multilayer
structures \cite{Baibich1988,Valet1993} of alternating
ferromagnetic and nonmagnetic layers and granular composites
\cite{Berkowitz1992,Xiao1992,Zhang1993,Milner1999} have been
studied widely in the last two decades. The giant
magnetoresistance (GMR) in these systems have been explained by
elastic scattering of the conduction electrons on magnetic moments
of differently aligned magnetic entities. Gittleman et al. have
shown that in superparamagnetic granular alloys this consideration
leads to a magnetoresistance proportional to the square of the
magnetization.\cite{Gittleman1972} Deviations from this relation
has been attributed to the size distribution of the magnetic
scatterers.\cite{Hickey1995,Ferrari1997} and interactions between
these scatterers \cite{Gregg,Kechrakos,Allia2003} Various
assumptions on the form of the size distribution have been made in
order to obtain a phenomenological description of the GMR
phenomena in different granular systems.

In this paper we present a systematical study of the magnetic and
magnetotransport properties of vacuum evaporated granular Fe-Ag
structures. The observed large, negative non-saturating magnetic
field dependence and the unusual sublinear temperature dependence
($d^2R/dT^2< 0 $) of the resistivity have been analyzed
simultaneously. This allowed the separation of the various
scattering processes and the identification of two characteristic
grain size determining the macroscopic magnetic and transport
properties without making any assumption on the grain size
distribution.

\section{EXPERIMENTAL}

The Fe-Ag multilayer samples were prepared by sequential vacuum
evaporation in a base pressure of $10^{-7}$ Pa onto Si(1\,1\,1)
single crystal substrates at room temperature. The mass of the
deposited material was measured by a quartz oscillator and the
nominal layer thickness was calculated using the bulk density of
Fe and Ag. In this paper
specimens prepared with the following sequences are discussed:\\
\\
(A) [Ag(2.6nm)/Fe(0.2nm)]$_{75}$/Ag(2.6nm)\\
(B) [Ag(1.3nm)/Fe(0.2nm)]$_{75}$/Ag(1.3nm)\\
(C) [Ag(0.8nm)/Fe(0.2nm)]$_{75}$/Ag(0.8nm)\\

Structural characterization of the samples by X-ray diffraction
indicated a nanometer scale grain size of the constituents [for
details see Ref. 13], however, due to the strong overlap between
the diffraction lines of bcc-Fe and fcc-Ag, a quantitative
evaluation of the size of the magnetic Fe grains was not possible.
The absence of peaks of the X-ray reflectivity\cite{Balogh2002}
when the Fe layer thickness is less than $\approx 1$ nm is
attributed to discontinuities of the Fe layers. This limit of the
continuous layer regime is quite similar to that observed in other
multilayer systems \cite{Rubinstein1994, Alphen1995,
Babenneau2000} composed of transition metals with immiscible
nonmagnetic elements and these kind of discontinuous multilayers
are often referred to as granular multilayers \cite{Hylton1993}.
The relation between the size of the grains in granular
multilayers and the nominal layer thickness is generally
determined by a three-dimensional growth process
\cite{Daruka1997}and the average diameter of the grains can be
much larger than the nominal layer thickness. It depends on the
material parameters (e.g. lattice parameter mismatch, surface
energy, etc.), as well as various parameters (deposition rate,
substrate temperature, etc.) of the deposition technique. The
decrease of the spacer layer thickness was found \cite{Balogh2005}
to increase the magnetic grain size in an other series of Fe/Ag
granular multilayer samples.

The magnetic structure of the samples was examined by a
superconducting quantum interference device (SQUID) and by
transmission M\"ossbauer spectroscopy. For the latter purpose the
samples have been removed from the substrate and folded up to
achieve an appropriate thickness for transmission measurements.

\begin{figure}[t!]
\includegraphics[width=\columnwidth]{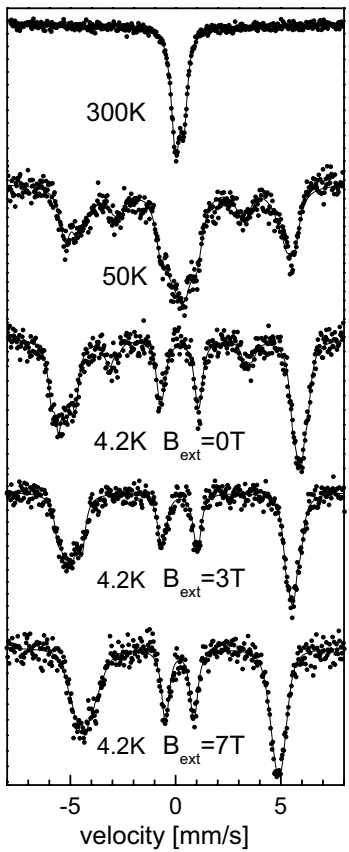}
\caption{Transmission M\"ossbauer spectra of sample A at various
temperatures. At $T=4.2$\,K spectra with and without an applied
magnetic field are shown.} \label{mossba.fig}
\end{figure}

\begin{figure}[t!]
\includegraphics[width=\columnwidth]{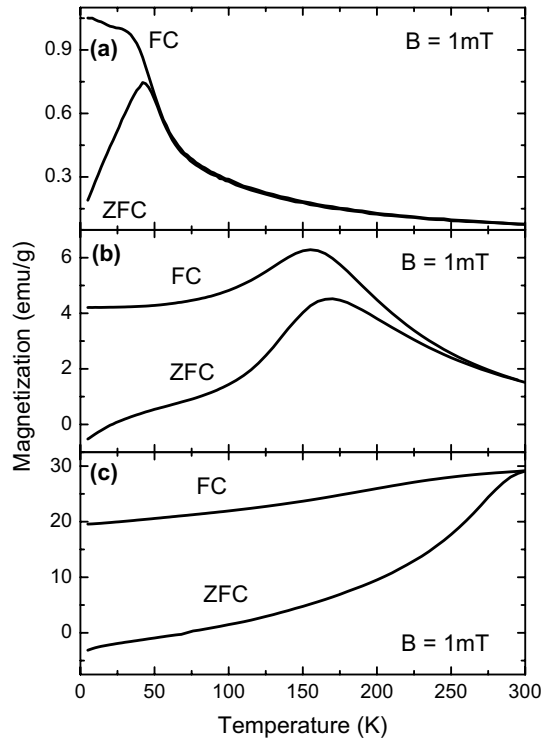}
\caption{Zero field cooled (ZFC) and field cooled (FC)
magnetization of samples A, B and C as a function of temperature
[panels (a), (b) and (c) respectively].} \label{susc.fig}
\end{figure}

The largest GMR and the strongest magnetic scattering of the
conduction electrons were observed in sample A, as it will be
shown later. The $^{57}$Fe M\"ossbauer spectra of this sample
taken at several temperatures are shown in Fig.~\ref{mossba.fig}.
The room temperature spectrum contains a paramagnetic doublet with
a large isomer shift relative to $\alpha$-Fe (0.18 mm/s) and a
quadrupole splitting (0.45 mm/s) characteristic to a system of
small Fe clusters embedded in an Ag matrix \cite{Sumiyama}. The
low temperature spectra show that the sample is superparamagnetic
(SPM) and as the magnetic clusters gradually freeze below the
blocking temperature (around 50K), the six-line pattern
characteristic to the Zeeman splitting of the nuclear levels of
$^{57}$Fe appears. At $T=4.2$\,K the paramagnetic fraction is
absent, but the spectral lines are much broader and the hyperfine
parameters are different than those of bulk bcc-Fe. The 4.2\,K
spectrum could be fitted with a distribution of hyperfine
fields\cite{Vincze}  with an average value, $H_{\rm av}=34.7$\,T
and standard deviation $\sigma_{\rm H}=$2.4\,T.

It is worth noting that the intensity ratios of the six lines
indicate a significant spontaneous alignment of the magnetic
moments. The intensity of the six lines of a sextet is distributed
as $3:I_{2-5}:1:1:I_{2-5}:3$, where $I_{2-5} = 4\sin2\Theta / (1 +
\cos2\Theta)$ is the intensity of the $m = 0$ transitions, and
$\Theta$ is the angle between the direction of the gamma-ray
(perpendicular to the sample plane) and the magnetic moment. In
case of a random distribution of the magnetization directions
$I_{2-5}=2$. The observed small intensity, $I_{2-5}= 0.5$,
indicates a close-to-perpendicular alignment of the magnetic
moments with respect to the sample plane. Applying a magnetic
field perpendicular to the sample plane could fully align the
moments parallel to the field, as it is indicated by the
$I_{2-5}=0$ intensity.

The spectra measured in external magnetic field were also fitted
with a distribution of the hyperfine fields and the parameters
obtained are: $H_{\rm av}=32.5$ and 27.8 and $\sigma_{\rm H}=2.3$
and 2.4\,T for $B_{\rm ext}=3$ and 7\,T. The external magnetic field
does not affect the width of the distribution indicating that it
is not due to relaxation of the magnetic moments. The broad
spectral lines of the observed sextet result from a distribution
of the Fe neighborhoods, which can be due both to the large number
of surface atoms in small grains and to non-equilibrium mixing of
the elements \cite{Burgler} during the growth process. The
field-independent width and the decrease of the observed average
hyperfine field in external field indicate the ferromagnetic
alignment of the magnetic grains. The hyperfine field of a
ferromagnet is decreased by the applied field, since it is
oriented antiparallel to the magnetic moment. We note that at 4.2\,K
the statistical errors allow an SPM fraction containing less
than 2 atomic \% of the Fe atoms.

The freezing of the superparamagnetic moments -- seen in the
M\"ossbauer spectra -- also appears in the temperature dependence
of the low field susceptibility measured by the SQUID.
Figure~\ref{susc.fig} shows the results for the 3 samples, after
cooling them either in zero or in 1 \,mT permanent magnetic field.
The blocking temperatures of about 40, 150 and 300\,K for samples A,
B and C, respectively can be deduced from this experiment. Sample
A ($T_{\rm B}= 40$\,K) exhibits a textbook example for the
superparamagnetic behavior. The maximum of the ZFC curve is more
smeared out for samples B and C and in case of sample B the
irreversibility temperature, where the FC curve starts to deviate
from the ZFC one, is much larger than the $T_{B}$ defined by the
ZFC maximum. Such features are generally
explained\cite{Chantrell,Binns} as the effect of a grain size
distribution and interaction among the magnetic particles, which
obviously play a role as the average grain size and the
concentration of the magnetic component increase.

\begin{figure}[b!]
\includegraphics[width=\columnwidth]{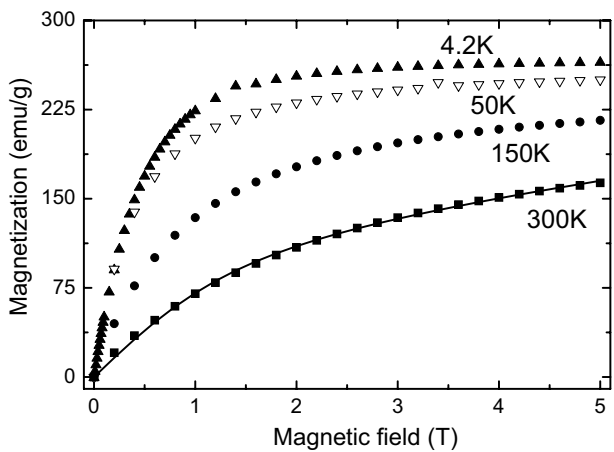}
\caption{Magnetization of sample A up to $B=5$\,T magnetic field at four
different temperatures. The solid line on the room temperature
curve confirmes the SPM Langevin model with a characteristic
magnetic moment of 500\,$\mu_{\rm B}$.} \label{squid.fig}
\end{figure}

The magnetization versus applied magnetic field curves of sample A
are shown in Fig.~\ref{squid.fig} up to $B=5$\,T at four different
temperatures. Above the blocking temperature the
magnetization can quite well be described\cite{Kiss2001} by a single Langevin
function, as expected for a superparamegnetic system of uniform
grains. The average magnetic moment of the SPM grains deduced from
the fit is $\mu_{\rm G}=$ 500\,$\mu_{\rm B}$. It indicates that the bulk
magnetic properties are mainly determined by the average size
grains (about 1.8\,nm) and the effect of the size distribution is
negligible.

\begin{figure}[t!]
\includegraphics[width=\columnwidth]{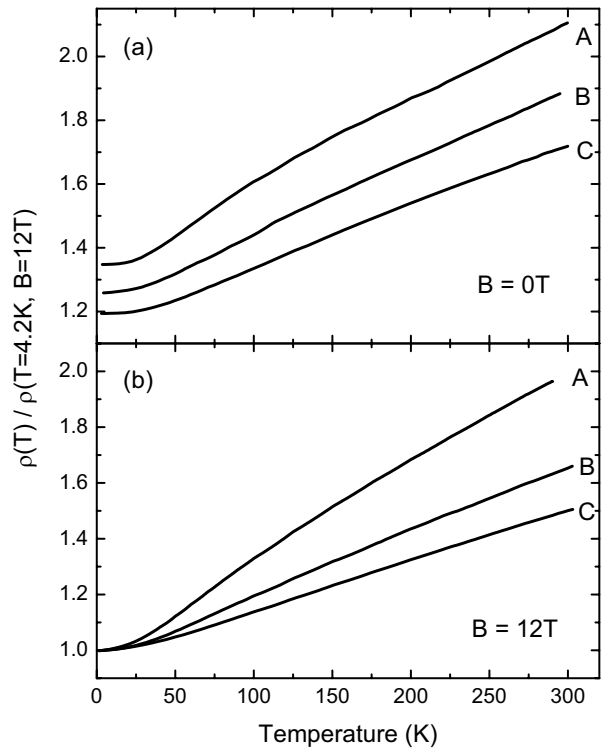}
\caption{Resistivity of samples A\,--\,C (normalized to their
$T=4.2$\,K, $B= 12$\,T values) as a function of temperature at zero
(a) and 12 T magnetic field (b).} \label{rt.fig}
\end{figure}

\begin{figure}[t!]
\includegraphics[width=\columnwidth]{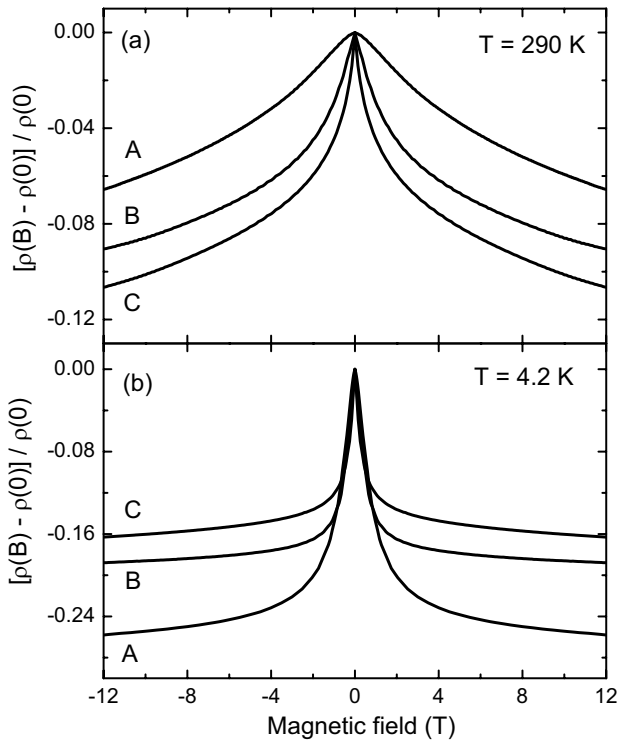}
\caption{Magnetoresistivity of samples A\,--\,C up to 12\,T magnetic
field at room temperature (a) and $T=4.2$\,K (b).} \label{gmr.fig}
\end{figure}

The magnetotransport measurements have been performed in the
current in plane arrangement by four contact method. The
coincidence of the parallel ($H\parallel I$) and transversal
($H\perp I$) magnetoresistance and the absence of any anisotropic
component is characteristic to the GMR phenomenon in granular
systems\cite{Xiao1992}. The temperature dependence of the
resistivity of samples A\,--\,C in zero and 12\,T magnetic field
normalized to their $T=290$\,K values are displayed in
Fig.~\ref{rt.fig}. Contrary to ordinary metallic systems, the
resistivity is sublinear above 40 K for each sample. Similar
observation was reported by Milner et~al. for granular
systems,\cite{Milner1999} but only in presence of a high magnetic
field. As discussed later in detail, the above qualitative feature
of the zero field temperature dependencies signifies the presence
of an extremely strong magnetic scattering in our samples.

The anomalous character of the $\rho(T)$ curves is the most
dominant in case of sample A. Simultaneously, this sample exhibits
the largest magnetoresistance at low temperature, as shown in the
bottom panel of Fig.~\ref{gmr.fig}. The magnitude of the GMR
measured at $T= 4.2$\,K in a field of $B= 12$\,T is 26\,\%, 18\,\% and
16\,\%, for samples A, B and C, respectively. Note that at room
temperature this order is reversed.

The magnetoresistance curves of the three samples have a common
feature, they do not saturate even in high magnetic fields and at
low temperatures, where the magnetization seems already to be
saturated. This indicates that significant magnetic scattering
takes place at magnetic entities much smaller than the typical
grain size which determines the macroscopic magnetization.  It is
well known that the scattering amplitude of ferromagnetic grains
embedded in a nonmagnetic metallic matrix is size dependent and
the contribution of the smaller clusters is strongly
enhanced.\cite{Zhang1993} The magnetic moments of these smaller
clusters are harder to rotate by an applied magnetic field thus
the saturation of the magnetoresistance is slower than that of the
net magnetization arising dominantly from the larger grains. Note
that similar behavior was found found in Fe-Ag co-deposited
granular films \cite{Makhlouf} as well as in many other systems
(for a review see \cite{Batlle}). In the next section we give a
more detailed analysis of the above qualitative picture.

\section{ANALYSIS}

In order to identify the resistivity contribution of the magnetic
scattering process we assume that the Matthiessen-rule can be
applied, i.e the resistivity is composed of 3 terms:
\begin{equation}
\rho(T,B)=\rho_{\rm 0}+\rho_{\rm ph}(T)+\rho_{\rm magn}(T,B)\mbox{
.} \label{totres}
\end{equation}
Here $\rho_{\rm 0}$ denotes the residual resistivity, $\rho_{\rm
ph}(T)$ is the contribution arising from the phonon scattering,
and the remaining part is attributed to the magnetic scattering.
The separation of the latter term requires further assumptions.

\begin{figure}[t!]
\includegraphics[width=\columnwidth]{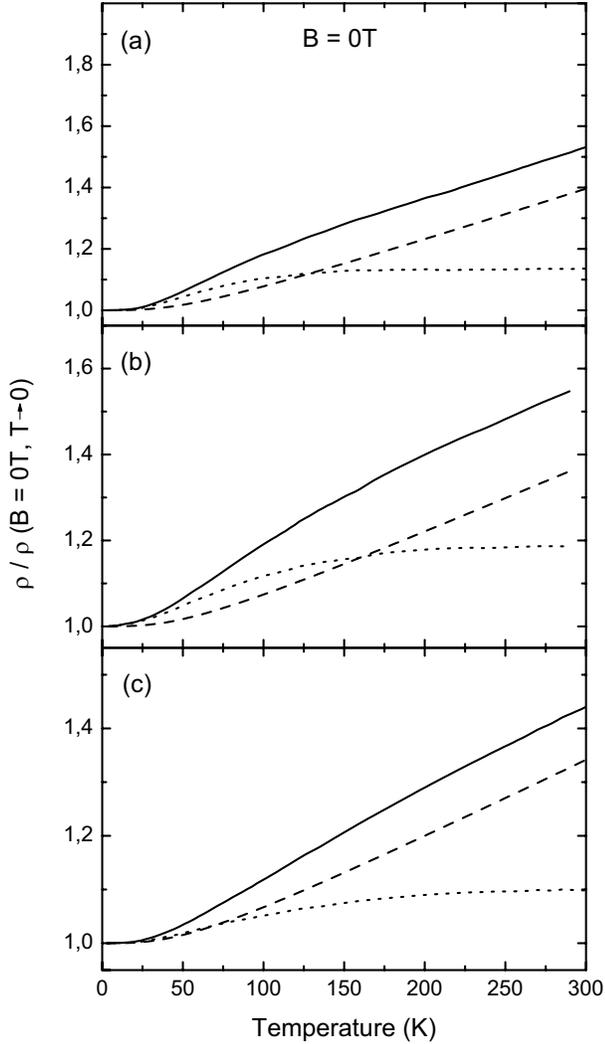}
\caption{Analysis of temperature dependence of the resistivity of
samples A\,--\,C [(a)\,--\,(c), respectively] at zero magnetic
field. The solid lines are experimental data, dashed lines
indicate the temperature-dependent resistivity contribution
arising from phonon scattering as calculated from the high
temperature slope of $\rho$(T) and the assumed Debye-temperature.
The dotted curves are the differences of the former two shifted
upwards by the residual resistivity, and are attributed to the
temperature-dependent magnetic scattering on small Fe clusters.}
\label{scatter0.fig}
\end{figure}

\begin{figure}[t!]
\includegraphics[width=\columnwidth]{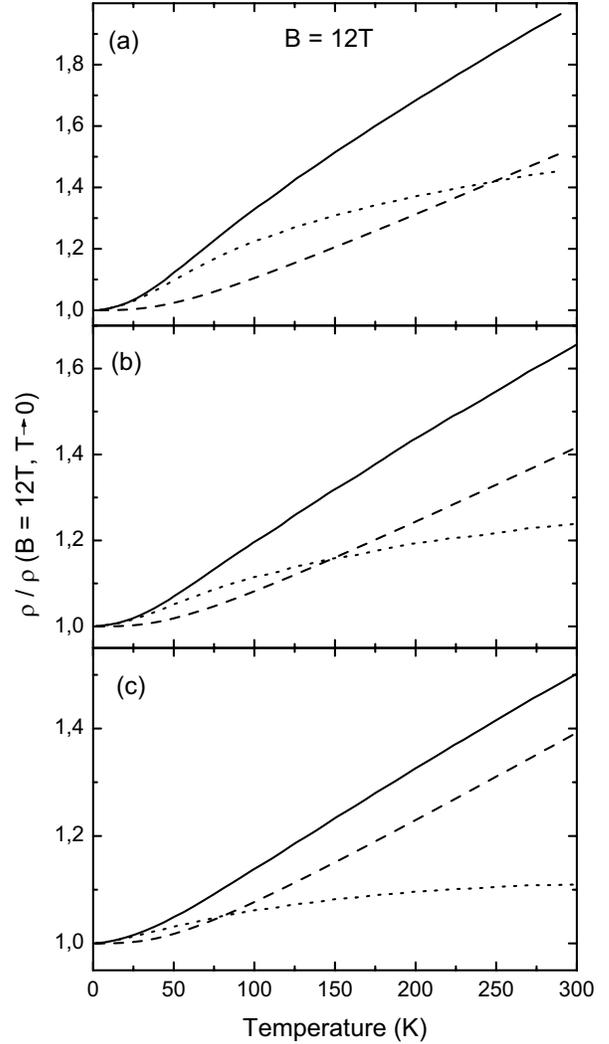}
\caption{Separation of the temperature-dependent resistivity
contributions related to different scattering mechanisms at $B=12$\,T
magnetic field. The magnetic field independent phonon part is
indicated by dashed line, the magnetic scattering related
contribution is plotted by dotted line for each sample [A\,--\,C
in panels (a)\,--\,(c), respectively].} \label{scatter12.fig}
\end{figure}

\begin{figure}[t!]
\includegraphics[width=\columnwidth]{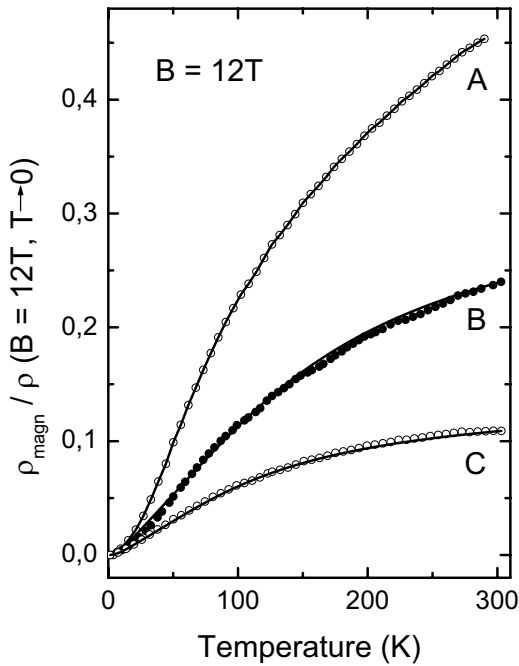}
\caption{Temperature dependence of the resistivity contribution
attributed to magnetic scattering on tiny Fe clusters in $B=12$\,T
magnetic field. Symbols are experimental data after subtracting
the phonon contribution and as shown in Fig.~\ref{scatter12.fig},
solid curves are calculated from Eq.~\ref{magn} with fitting
parameters $S= 16.6$ $\mu_{\rm B}$, 17.0 $\mu_{\rm B}$ and 12.5
$\mu_{\rm B}$ for samples A\,--\,C, respectively.}
\label{scatterfit.fig}
\end{figure}

\begin{figure}[t!]
\includegraphics[width=\columnwidth]{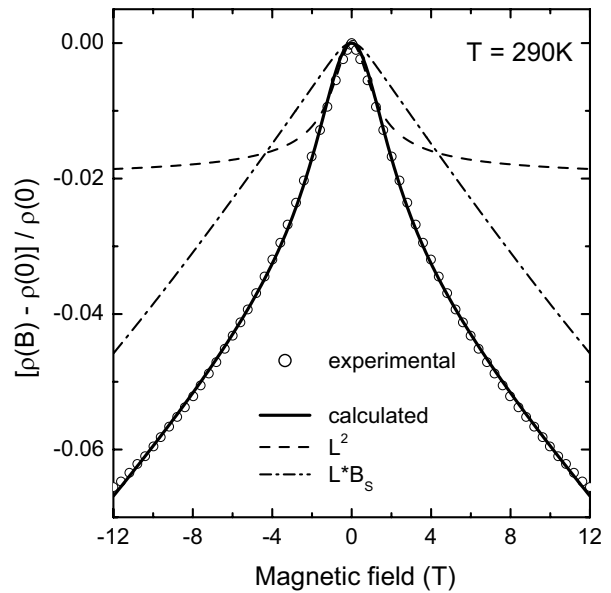}
\caption{Magnetoresistivity of sample A in the superparamagnetic
range, at room temperature. The open symbols are experimental
data, the solid line is the fit according to Eq.~\ref{chen}. The
dashed and dash-dotted curves correspond to the first and second
terms of Eq.~\ref{chen}, respectively.  The magnetic moments of
the clusters and the grains were taken from independent
experiments, see text for details.} \label{gmrfit.fig}
\end{figure}

Instead of making assumption on the shape of the grain size
distribution function\cite{Hickey1995,Ferrari1997} we use physical
considerations to get insight into the size distribution of the
grains determining the magnetic scattering. It is based on the
analysis of the temperature dependence of the resistivity measured
in zero and high magnetic field, as shown in
Figs.~\ref{scatter0.fig} and \ref{scatter12.fig}. In these limits
the large grains -- which dominate the bulk magnetic properties --
do not give temperature dependent contribution to the magnetic
scattering, as the system is either nonmagnetic (zero field limit,
above the blocking temperature) or fully polarized
ferromagnetically (high field limit). We will see that the small
clusters can also be characterized by a single average size, and
in high fields the temperature dependence of the magnetic
scattering is determined solely by this characteristic size: the
magnetic moment of these small clusters is the only fitting
parameter for the calculated curves describing the shapes of
$\rho_{\rm magn}(T,B=12 {\rm T})$ shown in Fig.~\ref{scatterfit.fig}.
Finally, the consistency of our analysis will also be demonstrated
by evaluating the magnetic field dependence of the resistivity
\emph{in the whole magnetic field range} (Fig.~\ref{gmrfit.fig})
using the characteristic sizes determined independently from the
magnetization data and the temperature dependence of the
resistivity in the zero and high field limits.

In a granular system the magnetic scattering depends on the
correlation between the localized magnetic moments of the grains,
$\left\langle {\vec{\mu}}_{\rm i} {\vec{\mu}}_{\rm
f}\right\rangle$, while a conduction electron is scattered from
${\vec{\mu}}_{\rm i}$ to ${\vec{\mu}}_{\rm f} $ within its
spin-diffusion length\cite{Gittleman1972}. In zero magnetic field,
well above the blocking temperature the magnetic moments of the
grains (including the large grains) are fully disordered. In these
circumstances no temperature dependence is expected from the
magnetic scattering, at least until the spin diffusion length is
large enough. We assume that at high temperatures the temperature
dependence of the resistivity arise solely from the phonon
contribution,
\begin{equation}
\rho_{ph}(T)=a_{1}\left(\frac{T}{\Theta}\right)^{3}\int_{0}^{\Theta/T}\frac{x^{2}dx}{e^{x}-1}\mbox{
.} \label{ph}
\end{equation}
As the the phonon term is linear above the Debye temperature
($\Theta \approx 210$ K)\cite{Milner1999}, the strength of phonon
scattering, $a_{1}$, can be determined from the high temperature
slope of the zero field resistivity curves. The calculated
$\rho_{\rm ph}(T)$ curves are shown in Fig.~\ref{scatter0.fig} by
dashed lines for the 3 samples. The difference of the total
resistivity and its phonon related part is attributed to the
magnetic scattering, and $\rho_{\rm 0}+\rho_{\rm magn}(T,0)$ is
displayed in Fig.~\ref{scatter0.fig} by dotted lines for each
sample. As it was expected, the magnetic scattering is temperature
independent at high temperatures and the lower the blocking
temperature the wider is the flat part of the separated magnetic
contribution curve. It is also seen that the magnetic scattering
gradually decreases as the blocking temperature is approached from
above. Note, however, that at
 $T=4.2$\,K there is still a considerable contribution from it,
 i.e. $\rho(T=4.2 {\rm K},B=0 {\rm T})\neq \rho (T=4.2{\rm K},B=12 {\rm T})$
 as it can be seen from Fig.~\ref{rt.fig}.

Since the phonon term is magnetic field independent, the
$\rho_{\rm ph}(T)$ curves determined from the zero field
temperature dependencies can be used to separate the magnetic
scattering contribution in the high field measurements. This is
shown in Fig.~\ref{scatter12.fig} for the $B=12$\,T measurements. A
comparison of the related panels of Fig.~\ref{scatter0.fig} and
Fig.~\ref{scatter12.fig} reveals that sample A has the strongest
negative curvature of the resistivity and the biggest change in
$\rho_{\rm magn}(T,B=12 {\rm T})$. In this sample magnetic
scattering at $B=12$\,T dominates over even the phonon term in a
very broad temperature range.

In order to describe the evaluated $\rho_{\rm magn}(T,B=12 {\rm
T})$ curves we assume that in the high field limit the magnetic
scattering of the spin-polarized electrons is proportional to the
spin disorder of the small clusters. The magnetic moments of the
large grains are fully aligned by the applied magnetic field, as
it could be deduced from the M\"ossbauer spectra of
Fig.~\ref{mossba.fig}. The spin disorder for a characteristic
moment $S$ is described by the Brillouin-function:
\begin{eqnarray}
\rho_{\rm magn}(T,B)&=&a_{2}\left(S-\left\langle S_{z}\right\rangle\right)\nonumber\\
&=&a_{2}\Bigg[S-\left(S+\frac{1}{2}\right)\coth\frac{(2S+1)g\mu_{B}B}{2kT}\nonumber\\
&+&\frac{1}{2}\coth\frac{g\mu_{B}B}{2kT}\Bigg]\mbox{ .}
\label{magn}
\end{eqnarray}
Here, $S$ and $\left\langle S_{z}\right\rangle$ are the total spin
and its $z$-component of the scatterers, respectively.

The fitted $\rho_{\rm magn}(T,B=12{\rm T})$ curves are shown in
Fig.~\ref{scatterfit.fig} by solid lines. Apart from a
normalization factor, the only fitting parameter is the magnetic
moment characteristic to the small Fe clusters. The good agreement
of the experimental and the calculated curves indicates that the
size distribution of these clusters is negligible. The fitted
values are in the same order of magnitude for all the samples;
$S=16.6$\,$\mu_{\rm B}$, 17\,$\mu_{\rm B}$ and 12.5\,$\mu_{\rm B}$ for samples
A\,--\,C respectively.

Next we discuss the magnetic field dependence of the resistivity
of sample A, which exhibits the strongest magnetic scattering in
the superparamagnetic state, where the process of magnetic
saturation is well understood. For a numerical analysis we use the
two characteristic sizes determined from the previous experiments.
The Langevin fit to the magnetization experiments performed in the
superparamagnetic temperature range has shown the presence of
large grains with $\mu_{\rm G}\approx 500$\,$\mu_{\rm B}$
(Fig.~\ref{squid.fig}), while the temperature dependence of the
resistivity in high magnetic field indicated the presence of small
clusters with $S\approx 17$\,$\mu_{B}$. Following Gittleman's
model \cite{Gittleman1972} we describe the magnetoresistance by
the field dependence of the correlation between the localized
magnetic moments responsible for an initial and a final magnetic
scattering process:
\begin{equation}
\frac{\Delta\rho}{\rho}\propto\left\langle\vec{\mu}_{i}\vec{\mu}_{f}\right\rangle
\propto\left\langle\vec{\mu}_{i}\vec{B}\right\rangle\left\langle\vec{\mu}_{f}\vec{B}\right\rangle\mbox{
.} \label{Git}
\end{equation}
In case of the observed two largely different grain sizes this can
be expressed as
\begin{eqnarray}
\frac{\Delta \rho}{\rho}&=& -b_{1}L^{2}\left(\frac{\mu
B}{kT}\right)-b_{2}L\left(\frac{\mu B}{kT}\right)B_{S}(B,T)\nonumber\\
&&-b_{3}B_{S}^{2}(B,T)\label{chen}
\end{eqnarray}
where $B_{\rm S}=\left\langle S_{\rm z}\right\rangle/S$ is the
Brillouin-function defined under Eq.~\ref{magn} for the small iron
clusters and $L(x)=\coth x-1/x$ is the Langevin-function, i.e. the
classical limit of $B_{\rm S}$ for the large Fe grains. The parameters
$b_1$, $b_2$ and $b_3$ represent the relative weights of
scattering from grain to grain, between a grain and a cluster and
from cluster to cluster.

Figure~\ref{gmrfit.fig} shows magnetoresistance for sample A in
the superparamagnetic phase and the expected variation calculated
by Eq.~\ref{chen}. The two characteristic magnetic moments
($\mu_{\rm G}=$ 500\,$\mu_{\rm B}$ and $S=$ 17\,$\mu_{\rm B}$ ) are determined in the
previous analysis from independent experiments: the field and
temperature dependence of the bulk magnetization, and the
temperature dependence of the resistivity. The relative weights of
the various processes were used as fitting parameters:
$b_1=0.035$, $b_2=0.013$ and $b_3=0.0001$. The value of $b_1$, i.e
the relatively weight for the grain$\rightarrow$grain scattering
process is larger than  $b_2$, even though the amplitude of the
grain scattering is small.\cite{Zhang1993} It reflects the large
probability of scattering from grain to grain due to the large
volume fraction of this type of magnetic scatterer. The direct
interplay between the clusters is negligible, as expected for a
small fraction (less than 2\,\%).

In Fig.~\ref{gmrfit.fig} the dashed and the dashed-dotted lines
represents the contribution of the two dominant scattering
processes. The grain$\rightarrow$grain scattering process is the
leading term in Eq.~\ref{chen} up to 4\,T magnetic field. The
dashed line corresponds to the square of the magnetization, and if
this would be the only scattering process the simple
relation\cite{Gittleman1972} of $\Delta R(B) \propto M^2$  would
hold. Above $B=4$\,T, however, the second term of Eq.~\ref{chen}
dominates and the observed behavior can be attributed to the
effect of the small clusters, not seen in the magnetization
curves. The good description of the measured magnetization and
resistance curves by two characteristic sizes is the consequence
of the narrow size distributions of the grains and the clusters.

Bimodal distribution of the grain size has already been observed
in granular systems prepared by co-deposition
\cite{Gregg,Franco,Pogorelov}, rapid quenching from the melt,
\cite{Hickey}, or layered growth \cite{Nouvertne,Venus} of the
constituents. Since for our samples a non-saturating
magnetoresistance indicates the presence of small clusters even in
case of 25 nm thick continuous Fe layers \cite{Balogh2002SSC}, we
associate the large grains and the small culsters to Fe rich
grains of the granular layers and small Fe clusters trapped inside
the Ag layers, respectively. Intermixing of the layers can occur
during the sample growth even when the heat of mixing is positive
\cite{Burgler}, like in the case of Fe and Ag. On the other hand,
the tendency for non-equilibrium mixing does not seem to depend on
the sample preparation method, since the magnetoresistance and the
magnetic properties of our granular multilayers are very similar
to those observed in co-depositied \cite{Makhlouf,Wang1994}
samples.

In our case for sample A the magnetic moments of the large grains
and the small clusters differ more than an order of magnitude;
$\mu_{\rm G}=$500\,$ \mu_{\rm B}$, and $S\approx$ 17\,$ \mu_{\rm B}$. According to the
M\"ossbauer spectroscopy measurements the small clusters contain
only a small fraction (below 2\,\%) of the magnetic atoms. This
explains why the bulk magnetization is determined by the
properties of the large grains at all temperatures, while in the
transport properties the magnetic scattering of small clusters
also play an important role.

In case of sample B and C a broader grain size distribution is
indicated by the smeared out ZFC curves and to describe the
magnetic field dependence of the resistance would require further
parameters. Interactions between the grains are also likely to
play a role \cite{Kechrakos} as the Fe concentration increases.
However, the temperature dependence of the resistivity measured in
12\,T magnetic field is quite similar to that observed in sample A
(see Fig. 4) and the analysis of the magnetic scattering
contribution (Fig. 8) undoubtedly indicates that there is a
significant contribution from small clusters in these samples, as
well.
\\

\section{SUMMARY}

In conclusion, we investigated the magnetic scattering processes
in sequentially evaporated granular Fe-Ag films. Unusual
magnetotransport features -- like sublinear temperature dependence
of the resistivity over a wide temperature range both in zero and
12\,T magnetic fields and large, non saturating GMR -- were found
experimentally. The contribution of the magnetic scattering was
separated and analyzed. The quantitative description suggests a
granular system with bimodal size distribution of the magnetic
components: coexisting large grains and small clusters. A detailed
numerical analysis was given to determine characteristic grain-
and cluster-moments, and their influence on both the scattering
processes and on the macroscopic magnetization. The analysis
reveals that scattering on the small clusters plays a dominant
role in high magnetic fields over a wide temperature range.

\section*{ACKNOWLEDGEMENTS}

Financial support of the Hungarian Research Founds OTKA TS049881,
T034602 and T038383 are acknowledged.

\end{document}